\documentclass[english,prb,aps,twocolumn,floatfix,superscriptaddress]{revtex4-1}

\usepackage[linktocpage,bookmarksopen,bookmarksnumbered]{hyperref}
\usepackage{graphicx}
\usepackage{dcolumn}
\usepackage{amsmath,graphics,epsfig,color,verbatim,ulem}
\usepackage{amssymb}
\usepackage{babel}

\begin{document}

\title{Influence of quantum confinement and strain on orbital polarization of four-layer LaNiO$_3$ superlattices: a DFT+DMFT study}

\author{Hyowon Park}
\affiliation{Department of Physics, University of Illinois at Chicago, Chicago, Illinois 60607, USA}
\affiliation{Materials Science Division, Argonne National Laboratory, Argonne, Illinois 60439, USA}
\author{Andrew J. Millis}
\affiliation{Department of Physics, Columbia University, New York, NY 10027, USA}
\author{Chris A. Marianetti}
\affiliation{Department of Applied Physics and Applied Mathematics, Columbia University, New York, NY 10027, USA}

\date{\today}

\begin{abstract}

Atomically precise superlattices involving  transition metal oxides provide a unique opportunity to engineer correlated electron physics using strain (modulated by choice of substate) and quantum confinement (controlled by layer thickness). Here we use the combination of density functional theory and dynamical mean field theory (DFT+DMFT) to study Ni E$_g$ $d$-orbital polarization in strained LaNiO$_3$/LaAlO$_3$ superlattices consisting of four layers of nominally metallic NiO$_2$ and four layers of insulating AlO$_2$ separated by LaO layers. The layer-resolved orbital polarization is calculated as a function of strain and analyzed in terms of structural, quantum confinement, and correlation effects.  The effect of strain is determined from the dependence of the results on the Ni-O bond-length ratio and the octahedral rotation angles; quantum confinement is studied by comparison to bulk calculations with similar degrees of strain; correlation effects are inferred by varying interaction parameters within our DFT+DMFT calculations.  The calculated dependence of orbital polarization on strain  in superlattices is qualitatively consistent with  recent X-ray absorption spectroscopy and resonant reflectometry data. However, interesting differences of detail are found between theory and experiment. Under tensile strain, the two inequivalent Ni ions display orbital polarization similar to that calculated for strained bulk LaNiO$_3$ and observed in experiment. Compressive strain produces a larger dependence of orbital polarization on Ni position and even the inner Ni layer exhibits orbital polarization different from that calculated for strained bulk LaNiO$_3$.
\end{abstract}

\maketitle

\section{Introduction}

Much of the interesting physics of transition metal oxides arises from the unusual properties of strongly interacting electrons in partly occupied transition metal $d$-shells. A key property of a partly filled $d$-shell is orbital polarization, the relative occupancy of different $d$-levels~\cite{RMP.70.1039}.  Interest in the possibility of using ``orbital engineering'' to control orbital polarization and thereby obtain desired electronic properties  continues to grow given the improving capability of synthesizing atomic-scale superlattices involving transition metal oxides~\cite{RMP.86.1189,Triscone:2011}. Superlattices allow for metastable structures with a range of lattice strain and many permutations of layerings, most of which could not be achieved by standard  bulk crystal growth methods. Much attention has focussed on superlattice systems based on rare earth nickelates following the theoretical prediction\cite{PRL.100.016404,PRL.103.016401} that complete orbital polarization of one of the Ni $d$-multiplets might be realized in superlattices composed of alternating layers of LaNiO$_3$ and an insulating spacer layer and  that the cuprate-like band structure implied by the complete orbital polarization might lead to high T$_c$ superconductivity in the superlattice. More advanced DFT+DMFT calculations later suggested that this scenario
will not be realized\cite{PRL.107.206804}, and this theoretical prediction is consistent with current experimental observations\cite{PRB.88.125124}.

Measuring orbital polarization is challenging, especially in artificially synthesized superlattices where the small volumes of typical samples mean that many types of experiments are not practicable. However  recent experimental progress in x-ray absorption and resonant reflectivity measurements have provided very interesting information \cite{PRB.88.125124,PRL.107.116805,EPL.96.57004}.  Connecting these experiments to theory to achieve a comphrensive understanding of the factors involved in controlling the orbital physics is an important task. 

Theoretical studies of orbital polarization in nickelate heterostructures and films have appeared. Methods used include model system calculations,  density functional theory (DFT), DFT+U, and the combination of DFT and the GW approximation. Most studies however have used the combination of density functional theory and dynamical mean field theory (DFT+DMFT). Effects that have been analyzed include  quantum confinement~\cite{PRL.100.016404,PRL.103.016401,PRB.82.235123,PRL.107.206804,PRB.88.195116}, strain~\cite{PRL.103.016401,PRB.82.134408,PRB.82.235123,PRL.107.206804,PRB.88.195116,PRB.90.045128},  local chemistry~\cite{PRB.82.134408}, and the consequences of charge doping~\cite{PRL.110.186402,PRL.114.026801}.   

The energy window used to define the correlated orbitals is an important issue in beyond DFT calculations such as the DFT+DMFT methodology\cite{PRB.90.235103}. The general consensus is that for the rare earth nickelates the correlated subspace should be defined in terms of atomic-like $d$-orbitals defined using Wannier or projector techniques applied to a wide energy range spanning at least the full $p-d$ manifold~\cite{PRL.107.206804}  (this is also the choice made in standard DFT+U implementations). As noted by Peil et al~\cite{PRB.90.045128}, if one wishes instead to define the correlated manifold is defined in terms of the near fermi-surface $p-d$ antibonding bands, the interaction parameters must be strongly renormalized.

A key finding of the published theoretical work is that the Hunds coupling acts to suppress orbital polarization down to a level substantially below the value predicted by standard DFT~\cite{PRL.107.206804,PRB.90.045128}. However, DFT+U calculations in bulk LuNiO$_3$ demonstrated that a Jahn-Teller distorted structure is only slightly higher in energy\cite{PRL.109.156402} than the bond disproportionated ground state, and a sufficiently large ($\gtrsim 4\%$) tensile (cubic-tetragonal with the in-plane bonds being longer) strain stabilizes the Jahn-Teller distorted structure\cite{PRB.91.195138}. The effect of more modest tensile strain on  the Ni-O bond-length ratio and the octahedral rotation was studied for tensile-strained bulk LaNiO$_3$~\cite{PRB.90.045128}, and the calculated orbital polarization was found to be in good agreement with the X-ray experiment~\cite{PRB.90.045128}. 

This paper is motivated by  recent tour-de-force measurements~\cite{EPL.96.57004,PRB.88.125124} of orbital polarization of the two inequivalent Ni sites (ie. inner and outer Ni-sites) in (LaNiO$_3$)$_4$/($R$XO$_3$)$_4$ superlattices comprised of alternating layers of four unit cells of LaNiO$_3$ and four unit cells of an insulating spacer layer $R$$X$O$_3$  with $R$=La,Dy,Gd and $X$=Al,Ga,Sc. By varying $R$ and $X$, the in-plane lattice constant could be adjusted to provide either tensile or compressive biaxial strain on the LaNiO$_3$ material while differences between orbital polarization of the Ni ion
adjacent to the $R$XO$_3$ and orbital polarization of the Ni ion surrounded by other Ni ions provides some insight into chemical and quantum confinement contributions. 

We build on the previously introduced theoretical techniques to ask the question: can a state of the art DFT+DMFT calculation based on a realistic crystal structure account for the essential features of the experiment? We study (LaNiO$_3$)$_4$/(LaAlO$_3$)$_4$ superlattices with four NiO$_2$ layers alternating with four AlO$_2$ layers. We incorporate the effects of strain (implemented in the experiment by changing the LaAlO$_3$ to  other wide gap perovskite insulators) by fixing the in-plane
lattice constant. We compare the DFT+DMFT calculations to pure DFT calculations and within the DFT+DMFT calculations consider different values of the correlation parameters. We estimate quantum confinement effects by comparing our results to those obtained on strained bulk LaNiO$_3$.

The rest of this paper is organized as follows. In Section ~\ref{Calculation} we present the specifics of the calculations. In Section ~\ref{Results:OP} we present our calculated orbital polarization. In Section ~\ref{Analysis} we discuss the impact of structure on orbital polarization and in Section~\ref{manybody} we delineate the consequences of the many-body interactions.  Section ~\ref{Conclusion} is conclusion.

\section{Model and methods\label{Calculation}}

We study superlattices consisting of four layers of LaNiO$_3$ alternating with four layers of the wide bandgap insulator LaAlO$_3$ with the alteration along the (001) direction of the ideal cubic perovskite structure; we refer to the resulting structures as  (001) (LaNiO$_3$)$_4$/(LaAlO$_3$)$_4$  superlattices. We impose tetragonal symmetry, meaning that the two in-plane lattice constants $a$ are equal and the lattice vectors  are at right angles to each other and to the out-of-plane lattice vector. We allow for rotations and tilts of the NiO$_6$ and AlO$_6$ octahedra; this doubles the unit cell in-plane though the combination of an in-plane translation and a rotation maps one NiO$_6$ octahedron onto the other.  The four LaNiO$_3$ layers come in two equivalent pairs. We denote the Ni in the outer (closer to Al) layer as Ni B and the Ni in the inner layer as Ni A.

We simulate strain by varying the in-plane lattice constant and define tensile (compressive) strain as in-plane lattice constant larger (smaller) than the mean Ni-Ni distance $a_0$ calculated for bulk LaNiO$_3$; quantitatively,  (in-plane) strain is $(a-a_0)/a_0$.  The theoretical $a_0$ value obtained by performing a structural relaxation within the  GGA methodology is 3.87\AA $\:$ and the measured equilibrium volume used to define strain in the experiment is 3.838\AA~\cite{PRB.88.125124}.

In the first step in our calculations we use DFT to relax all internal coordinates and the (001) axis lattice  parameter under the assumption of tetragonal symmetry and fixed strain, meaning that the  in-plane lattice constants are fixed at definite values and constrained to be at right angles. (We thus neglect the small monoclinic distortion occurring in  bulk LaNiO$_3$; the effects of this distortion are considered in Ref.$\:$\onlinecite{PRB.90.045128}.) The structural relaxation of atomic positions is performed using the Vienna Ab-initio Simulation Package (VASP)~\cite{Kresse199611169,Kresse19991758}, a plane-wave DFT code based on the projector augmented wave formalism~\cite{Blochl:1994}.  The exchange-correlation potential is taken to be the spin-polarized generalized gradient approximation (s-GGA) using the Perdew-Burke-Ernzerhof (PBE) functional~\cite{PRL.77.3865} and in our calculations the ground state is ferromagnetic. The structural relaxation is converged if the atomic forces of all atoms are smaller than 0.01eV/\AA. We note that for physically relevant $U\gtrsim 5eV$, DFT+U relaxation calculations based on spin-polarized GGA or non-spin-polarized GGA  wrongly produce the bond disproportionated structure for LaNiO$_3$ even at  ambient pressure~\cite{PRB.89.245133,PRB.90.235103}. A $k-$point mesh of $8\times8\times1$ is used and the plane-wave energy cutoff $E_{cut}$ is set to be 600eV. Note that only one $k_z$ point is sufficient as the supercell is enormously elongated in the $z-$direction; we confirmed (not shown)  that a $8\times8\times2$ $k-$mesh relaxes to an essentially identical  structure ($\sim 0.1\%$ changes to octahedral $l_c/l_a$ ratio (see definition in Fig.$\:$\ref{fig:bondlength} caption) and $\sim 1\%$  changes to rotation angles). 
 
We specify the rotation patterns for given structure using Glazer notation\cite{Glazer:a09401,Glazer:a11802}. Experimentally, bulk LaNiO$_3$ has a rhombohedral unit cell (space group $R\bar{3}c$, Ref.$\:$\onlinecite{PRB.46.4414}) with the NiO$_6$ octahedral rotation of the $a^-a^-a^-$ pattern. The $a^-a^-c^+$ pattern is also observed for nickelates with a smaller rare-earth ion than La~\cite{PRB.64.094102}.  In our calculations, for strained bulk LaNiO$_3$ on a cubic substrate we find the $a^-a^-c^-$ pattern. Our strained LaNiO$_3$ superlattice shows asymmetric behavior depending on the sign of strain. Under compressive strain the relaxed structure exhibits the $a^-a^-c^-$ pattern of the NiO$_6$ octahedral rotation similar to strained bulk while the $a^-a^-c^+$ pattern becomes stable under tensile strain. 
%For the rotation patterns, we consider either $a^-a^-c^-$ or $a^-a^-c^+$ in Glazer notation~\cite{Glazer:a09401,Glazer:a11802} and they show asymmetric behavior depending on the sign of strain.

For each relaxed structure we use the generalized gradient approximation plus dynamical mean field theory (GGA+DMFT) to calculate the electronic structure. In these calculations we use the spin-unpolarized form of the GGA-PBE functional and constrain the DMFT calculations to the paramagnetic phase; this is appropriate since no magnetism has been observed in the system of experimental interest~\cite{Boris20052011,PRL.111.106804,PRB.88.125124}. The correlated subspace is taken to be the atomic-like Ni  $d$-orbitals defined by a standard maximally localized Wannier function construction~\cite{Wannier} based on a wide energy window spanning the full 12eV energy range of the Ni-$3d$ O-$2p$ band complex.  We follow
previous work ~\cite{PRL.109.156402,PRB.89.245133,PRB.90.235103} and rotate the orbital quantization axis on each Ni site to the direction  that minimizes the off diagonal components of the hybridization function. This direction is approximately aligned to the local Ni-O octahedral axes. The filled $t_{2g}$ orbitals are treated using a Hartree-Fock approximation while the full dynamics of the $e_g$ orbitals is considered, as in our previous studies~\cite{PRL.109.156402,PRB.89.245133,PRB.90.235103}.

We adopt  rotationally invariant Slater-Kanamori interactions within the Ni $d$ correlated subspace. The interactions are parametrized by a Coulomb repulsion $U$, a Hunds coupling $J$ and a double counting correction $U^\prime$. Our previous work ~\cite{PRB.89.245133,PRB.90.235103} showed that $U$=5eV, $U^\prime=4.8$eV and $J$=1eV provided the best fit to the structural and metal insulator phase diagram across the rare-earth nickelate family. In this paper we use these values but consider $J=0.7$eV as well as $J=1.0$eV to  provide insight into the $J$-dependence, since the value of $J$ was previously shown to be important for the quantitative value of the orbital polarization.~\cite{PRL.107.206804,PRB.90.045128}

The main quantity of interest in this paper is orbital polarization, $P$. A precise definition  is required, since the values obtained depend on the way the $d$-orbitals and polarization are defined. Here we adopt the defintion of $P$ used in Ref.$\:$\onlinecite{PRB.88.125124}, with the $d$-orbitals defined as the rotated Wannier orbitals $|W^a_{e_g}\rangle$ discussed above and $a=3z^2-r^2$ or $x^2-y^2$ denoting the $e_g$ orbitals of main interest here. The orbital occupancies are then the diagonal elements of the occupancy matrix

\begin{equation}
n_{e_g}^{aa} = \frac{T}{N_{\mathbf{k}}}\sum_{i\omega_n,\mathbf{k}}\sum_{(l,l')\in w} \langle W^a_{e_g}|\psi_{\mathbf{k}l}\rangle G^{\mathbf{k}}_{ll'}(i\omega_n)\langle\psi_{\mathbf{k}l'}|W^a_{e_g}\rangle,
\label{eq:n_wan}
\end{equation}
\begin{equation}
\hat{G}^{\mathbf{k}}(i\omega_n) = \frac{1}{(i\omega_n+\mu)\hat{\mathbb{I}}-\hat{H}^{KS}_{\mathbf{k}}+\hat{P}_{cor}^{\dagger}\cdot\hat{\Sigma}_d(i\omega_n)\cdot\hat{P}_{cor}}
\end{equation}
where $\hat{H}^{KS}_{\mathbf{k}}$ is the Kohn-Sham Hamiltonian at the $\mathbf{k}$ point in the Brillouin zone, $\psi_{\mathbf{k}l}$ is the corresponding Kohn-Sham eigenfunction with  band index $l$, $\Sigma_{d}$ is the self energy for the Wannier $d$ orbitals, and $\hat{P}_{cor}$ is a projection operator defined to downfold to the correlated $d$ subspace.
$\omega_n$ is the Matsubara frequency and $T$ is the temperature. Within GGA, $n^{aa}_{e_g}$  is obtained by inserting $\Sigma_{d}=0$.

Orbital polarization $P$ is then defined in terms of the hole density per spin $h^a=1-n^{aa}_{eg}$ as
\begin{eqnarray}
P=\left(\frac{4}{n^{atomic}_{e_g}}-1\right)\frac{(X-1)}{(X+1)}
\label{eq:pol}
\end{eqnarray}
with $X=h^{3z^2-r^2}/h^{x^2-y^2}$ and $n_{e_g}^{atomic}$ is the occupation value for the atomic-like Wannier function which we set  to  1.0 for consistency with  Ref.~\onlinecite{PRB.88.125124}. Thus positive $P$ means the lower relative occupancy of the $3z^2-r^2$ orbital.

\section{Results: Orbital Polarization\label{Results:OP}}

\begin{figure}[!htbp]
\includegraphics[width=0.85\columnwidth]{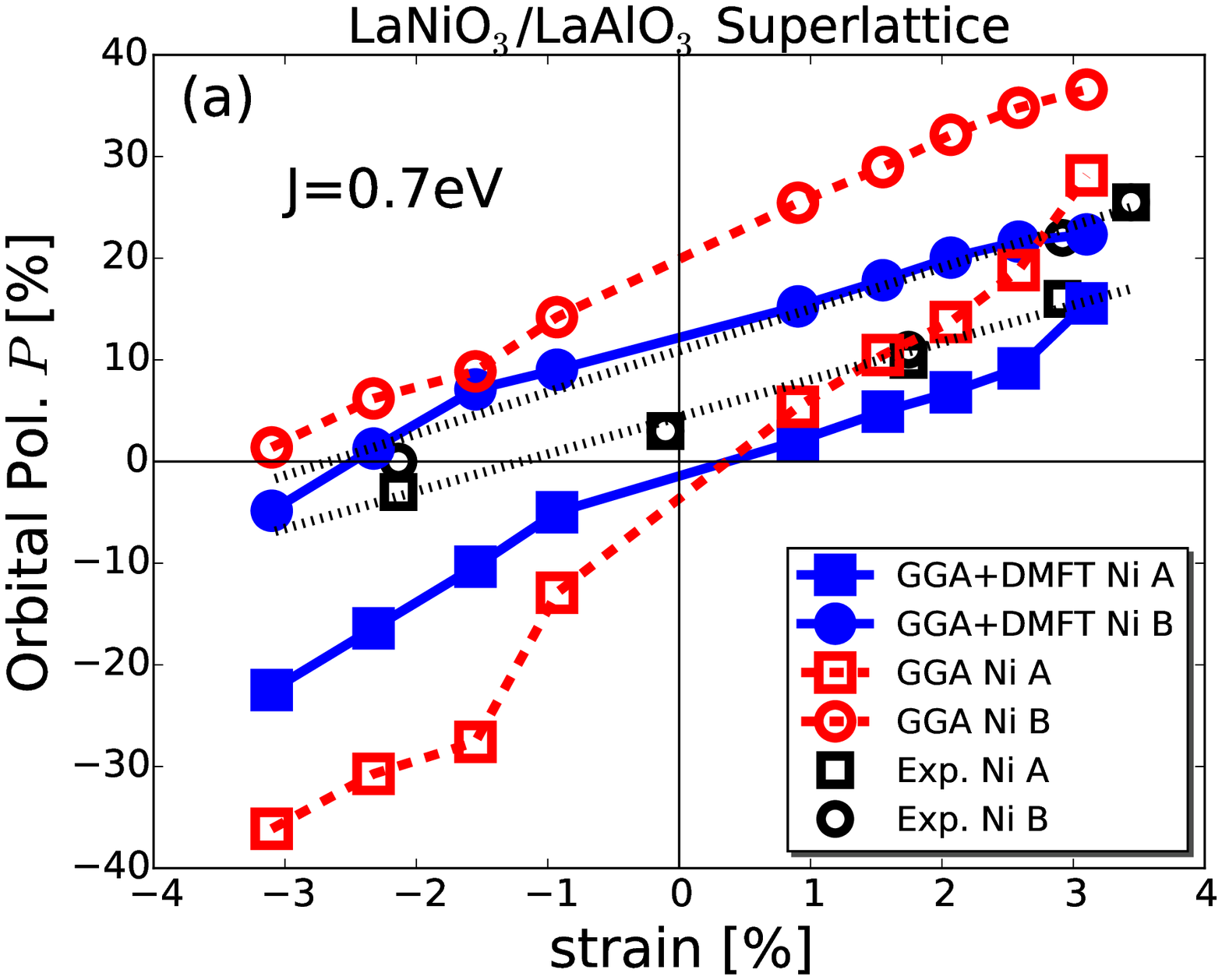}
\includegraphics[width=0.85\columnwidth]{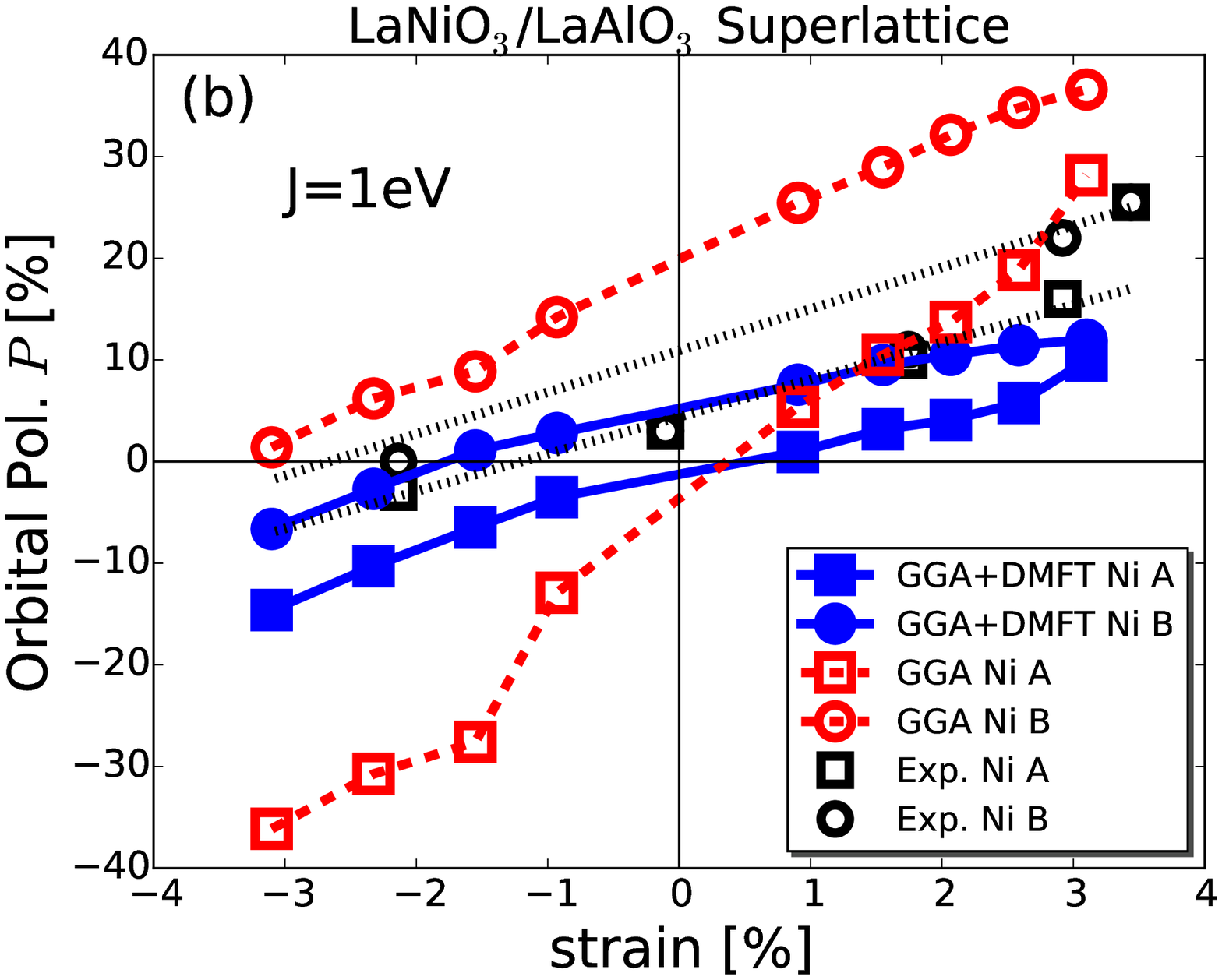}
\includegraphics[width=0.85\columnwidth]{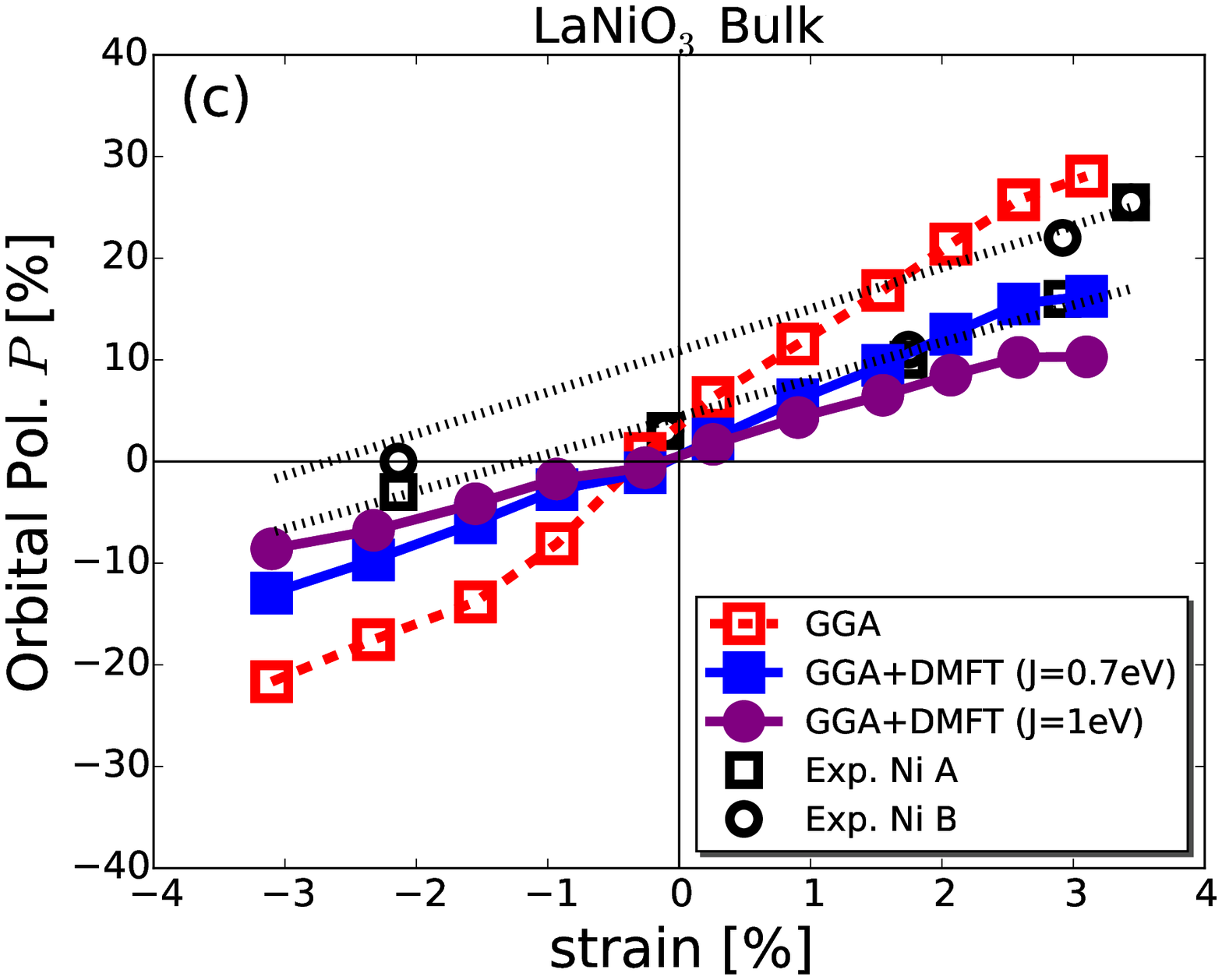}
\caption{(Color online) 
Orbital polarization $P$ (Eq.~\ref{eq:pol})  as a function of bi-axial strain for the two inequivalent Ni sites of the  4/4 LaNiO$_3$/LaAlO$_3$ superlattice (panels a and b) and for strained bulk LaNiO$_3$ (panel c)  computed using paramagnetic GGA+DMFT.   Interaction parameters of $U$=5eV and $J$=0.7eV (panel a),  $U$=5eV and $J$=1eV (panel b), and  $U=5$ eV and $J$=0.7/1.0eV (panel c) are used. Paramagnetic pure GGA results are also presented, as are experimental data (black empty dots, dashed lines) obtained from Ref.~\onlinecite{PRB.88.125124}.
\label{fig:pol}}
\end{figure}

Fig.$\:$\ref{fig:pol} shows our main results: orbital polarization $P$ computed for the superlattice and for bulk LaNiO$_3$  and compared to experimental data. Our results for bulk LaNiO$_3$ (panel c) are very similar to those presented by Peil $et$ $al$ ~\cite{PRB.90.045128}. 
The GGA results agree almost exactly under both tensile and compressive strain. The GGA+DMFT results are consistent given the difference in parameters (Peil $et$ $al$ considered $U$=8eV and $J$=1eV whereas we consider $U$=5eV and $J$=0.7eV; the increase in polarization due to the increase in $U$ is almost compensated by the decrease in polarization due to the increase in $J$. 
%; agreeing almost exactly for GGA under both tensile and compressive strain and, qualitatively for GGA+DMFT under tensile strain. 
%{\bf I believe Peil used J=1 and our J=1 results agree with his. Can you confirm?}

We see that both the GGA and the GGA+DMFT calculations for superlattices (panel a and b) are in reasonable qualitative correspondence with experiment both in terms of the order of magnitude of the change over the interesting strain range and in terms of the sign of the difference between polarizations of the  inner and outer Ni layers. Consistent with experiment, all calculations indicate that tensile strain increases orbital polarization (relative occupancy of $d_{3z^2-r^2}$ orbital) while compressive strain decreases it. Also consistent with experiment, the calculations indicate that at any value of the strain orbital polarization of the A (inner layer) site is less than that of the B (outer layer) site, meaning that the inner layer Ni ion has lower occupancy of the $x^2-y^2$ and higher occupancy of the $3z^2-r^2$ orbital than does the outer layer Ni ion. 

The observed change of $P$ with strain in the bulk material (Fig.$\:$\ref{fig:pol}c) may be qualitatively understood as a consequence of the antibonding nature of the near fermi surface bands. Considering for example tensile strain, the increase in $l_a$ reduces the hybridization of the planar Ni orbital to the surrounding oxygens, thus lowering the energy of  the $x^2-y^2$ derived band, decreasing the density of $x^2-y^2$ holes, while the concomitant decrease in $l_c$ conversely increases the hybridization to the $3z^2-r^2$ orbital, increasing the density of $3z^2-r^2$ holes. 

While all of the calculations are qualitatively consistent, interesting quantitative differences occur. The GGA calculations predict a stronger  dependence of orbital polarization on strain and on Ni site, and predict a stronger change across zero strain, than do the GGA+DMFT calculations, with the difference between GGA and GGA+DMFT being larger for larger $J$. This is an example of the physics discussed in Ref.~\onlinecite{PRL.107.206804}: the electronic configuration of the Ni atoms is closer to $d^8{\bar L}$ than to $d^7$ so as Hunds coupling is increased the probability that the Ni is in the high spin $d^8$ state increases, and in the high-spin $d^8$ state both $e_g$ orbitals are occupied so orbital polarization is suppressed.  However, as we shall show below the  details of the $J$-dependence of different aspects of the strain dependence is somewhat unexpected. 

\section{Analysis: structural effects \label{Analysis}}

\begin{figure}[t]
\includegraphics[width=0.85\columnwidth]{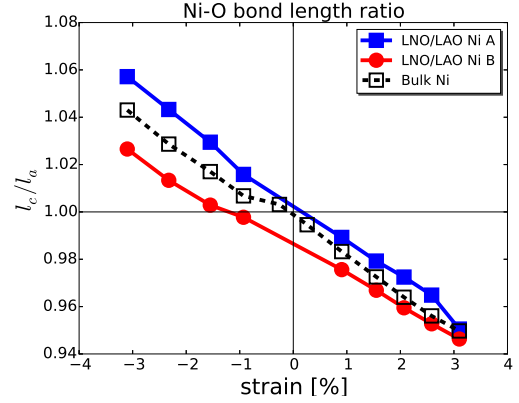}
\caption{(Color online) 
The layer-resolved Ni-O bond length ratio $l_c/l_a$ ($l_c$: out-of-plane bond length, averaged over the two out of plane bonds for a given Ni, $l_a$: in-plane bond length, averaged over the four in plane bonds for a given Ni) for Ni A and Ni B sites as a function of strain for 4/4 LaNiO$_3$/LaAlO$_3$ superlattice (filled dots, solid line) and bulk LaNiO$_3$ (empty dots, dashed line) 
\label{fig:bondlength}}
\end{figure}

In this section we analyze the relation between the structural distortions induced by strain and the orbital polarization. We begin with Fig.$\:$\ref{fig:bondlength}  which shows the calculated strain dependence of the octahedral distortion $l_c/l_a$ (ratio of apical to in-plane Ni-O bond lengths). The dashed line with open symbols  shows that for bulk LaNiO$_3$, as expected, an increase in the planar bond length (tensile strain) leads to a decrease in the c-axis bond length, and conversely. The slope of the $l_c/l_a$ curve implies that the bulk material has a calculated Poisson ratio $\nu=-(\delta l_c/\delta l_a)/(2-\delta l_c/\delta l_a)$  of roughly $0.25$ ($\delta l_c/\delta l_a=\delta(l_c/l_a)/\delta(\epsilon_x)+1$ where $\epsilon_x$ is the strain in the $x$-direction and note that the octahedral bond lengths do not correspond exactly to changes in lattice constants because the octahedral rotations also vary).  

\begin{figure}[b]
\includegraphics[width=0.85\columnwidth]{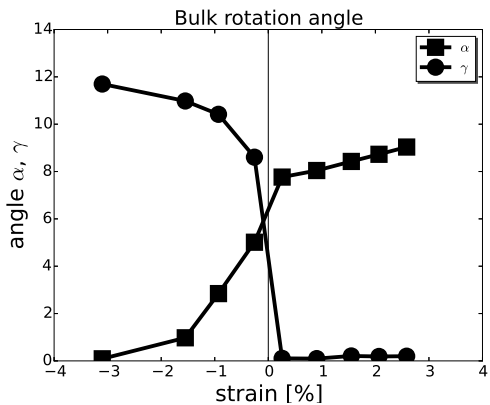}
\caption{(Color online) 
Strain dependence of octahedral rotation angles $\gamma$ (about axis normal to the stress plane) and $\alpha$ (about axis lying in the stress plane) for bulk LaNiO$_3$.
\label{fig:NiOAnglebulk}}
\end{figure}

A small anomaly in the $l_c/l_a$ ratio occurs at slightly compressive strain (-1\%).  This arises from an abrupt change in the octahedral rotation angle occurring at this strain, shown in Fig.~\ref{fig:NiOAnglebulk}.  Near zero strain, the strain effect is accommodated by the octahedral rotation rather than the octahedral distortion. The orientations are specified by rotation angle $\gamma$ about the axis normal to the plane which is stressed and $\alpha$ about an axis lying in the stress plane.  Consistent with experimental data\cite{PRB.82.014110}, for compressive strain the calculations indicate the dominant rotation is around the $z$ axis and the rotation about the in-plane axis is small ($\gamma \gg  \alpha \sim 0$),  while for tensile strain the rotation about the in-plane axis is large and there is  no rotation about the axis normal to the stress plane.

\begin{figure}[b]
\includegraphics[width=0.85\columnwidth]{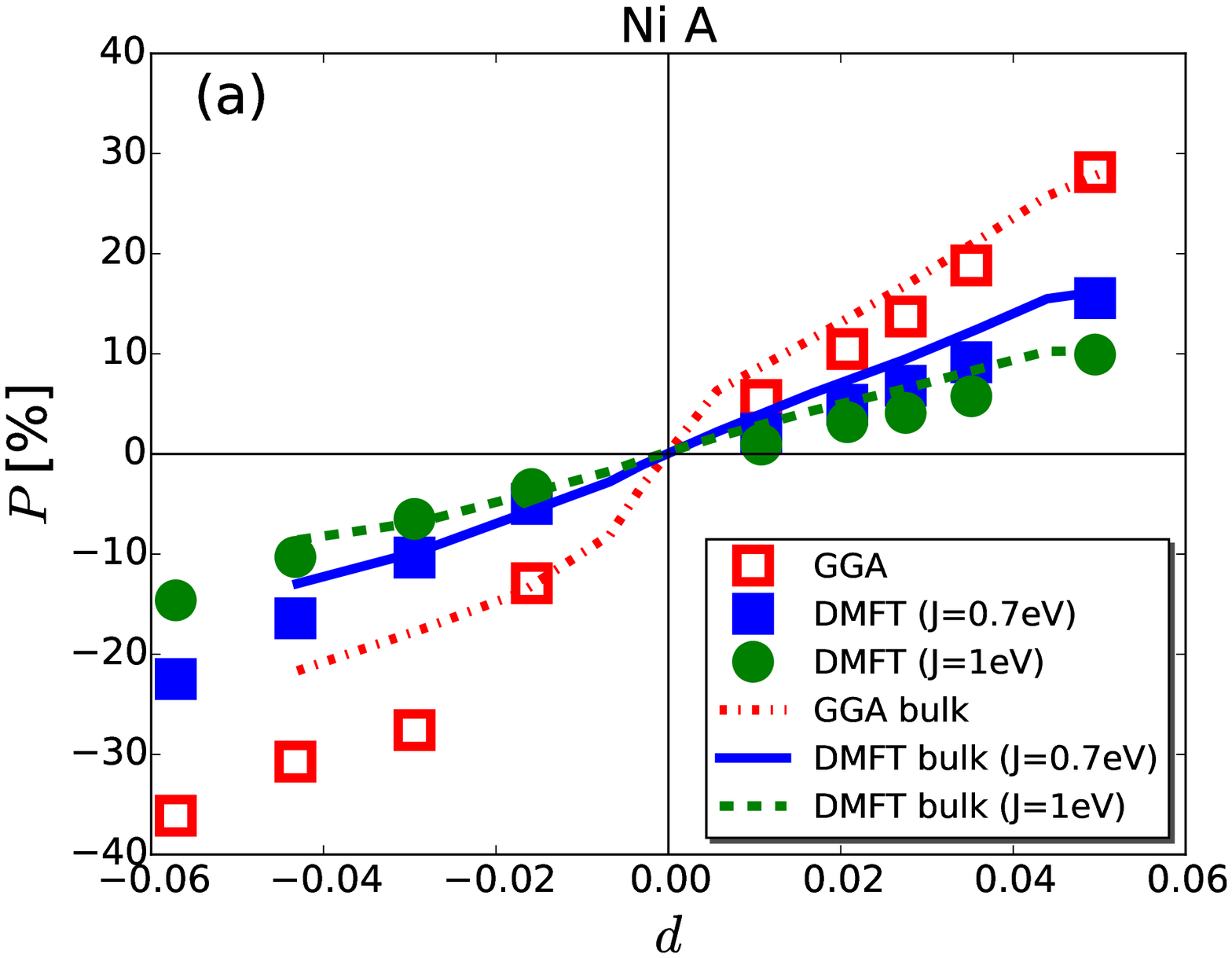}

\vspace{0.1in}

\includegraphics[width=0.85\columnwidth]{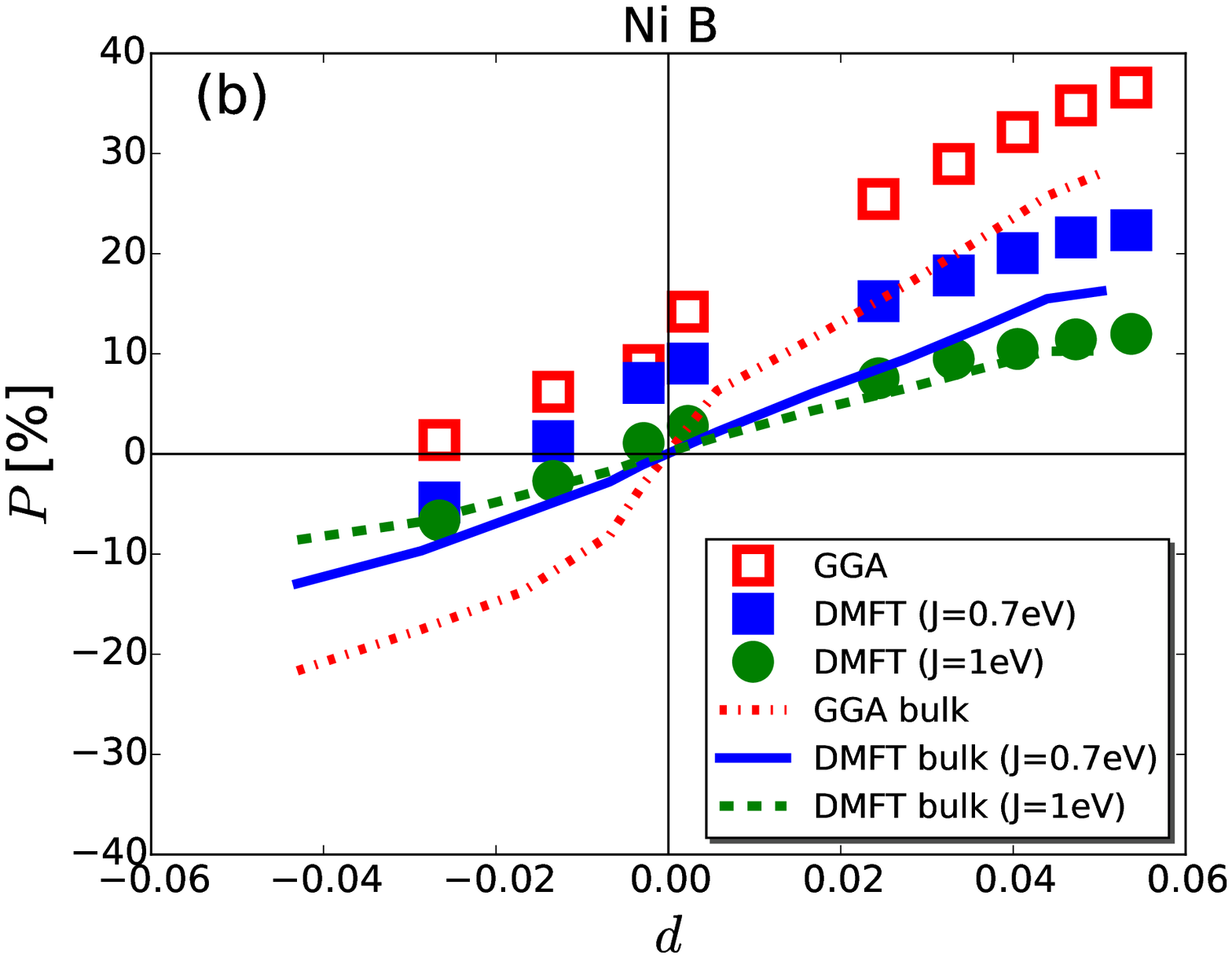}
\caption{(Color online) 
Orbital polarization as a function of octahedral distortion $d=\frac{l_c}{l_a}-1$ computed using GGA and GGA+DMFT (J=0.7 and 1.0 eV) for the (a) Ni A and (b) Ni B sites of the 4/4 LaNiO$_3$/LaAlO$_3$ superlattice (points) compared to results for strained bulk LaNiO$_3$ (lines). \label{fig:pollcla}}
\end{figure}

To obtain a more quantitative and precise understanding of the relation between structure and polarization we replot the data in Fig.~\ref{fig:pol}c in terms of octahedral distortion $d$ defined as
\begin{equation}
d=1-\frac{l_c}{l_a}
\label{distortiondef}
\end{equation}
Results are shown as dashed-dot (GGA), solid (DMFT $J$=0.7eV), and dashed (DMFT $J$=1eV) lines in Fig.~\ref{fig:pollcla}. We find that the dependence of $P$ on distortion is approximately linear and may be written as
\begin{equation}
P(d)=P_0+Rd
\label{Pofd}
\end{equation}
The slope $R$ defines the response of orbital polarization to an $E_g$-symmetry octahedral distortion, in other words the degree to which a distortion of a Ni-O octahedron leads to a differential occupancy of the Ni $e_g$ levels.  The intercept $P_0$ gives a measure of the other contributions to the orbital polarization; in the bulk case, the only other contribution comes from the octahedral rotations, while in the superlattices quantum confinement effects may also play a role.

Results for $P_0$ and $R$ are given in Table~\ref{linearfit}. We see that for the bulk material, the difference of $P_0$ values between between positive and negative $d$ is small, of the order of $10-20\%$ of the total change in $P$ across the strain range we study, and the difference in $R$ is also not large. However, the slope $R$ is strongly increased across the unstrained point and the $P_0$ value is almost zero (Fig.~\ref{fig:pollcla}). We interpret these changes as arising from the abrupt change in octahedral rotations across this point.  In agreement with previous work\cite{PRL.107.206804,PRB.90.045128}, we find that many-body effects, in particular increasing the Hunds coupling $J$, act to decrease both the magnitude of $P$ and its response to structural change $R$. It is interesting to note that the interactions decrease $P_0$ by a relatively larger amount than they reduce $R$ and that the renormalization of $R$ is larger for tensile strain.  We will discuss the reason in section ~\ref{manybody}.

\begin{table}
\begin{center}
  \begin{tabular}{@{} ccc @{}}
    \hline
  Bulk & Compressive  & Tensile  \\ 
    \hline
    $P_o$ GGA & -6.6 & 2.7 \\ 
   $P_o$ DMFT (J=0.7 eV) & -1.1 & 0.12 \\ 
   $P_o$ DMFT (J=1.0 eV) & -0.75 & 0.61 \\ 
   \hline
    $R$ GGA & 3.6 & 5.2 \\ 
    $R$ DMFT (J=0.7 eV) & 2.8 & 3.5 \\ 
    $R$ DMFT (J=1.0  eV)& 1.9 & 2.2 \\ 
    \hline \hline
  \end{tabular}

\vspace{0.1in}

 \begin{tabular}{@{} ccc @{}}
    \hline
  Ni A Site & Compressive  & Tensile  \\ 
    \hline
    $P_o$ GGA & -18 & -0.71 \\ 
   $P_o$ DMFT (J=0.7 eV) & 2.6 & -1.1 \\ 
   $P_o$ DMFT (J=1.0 eV) & 2.2 & -1.1 \\ 
   \hline
    $R$ GGA & 3.1 & 5.4 \\ 
    $R$ DMFT (J=0.7 eV) & 4.4 & 2.9 \\ 
    $R$ DMFT (J=1.0  eV)& 2.9 & 1.9 \\ 
    \hline \hline
  \end{tabular}

\vspace{0.1in}

  \begin{tabular}{@{} ccc @{}}
    \hline
  Ni B Site & Compressive  & Tensile  \\ 
    \hline
    $P_o$ GGA & 12 & 15 \\ 
   $P_o$ DMFT (J=0.7 eV) & 8.1 & 8.5 \\ 
   $P_o$ DMFT (J=1.0 eV) & 2.0 & 3.7 \\ 
   \hline
    $R$ GGA & 4.1 & 4.1 \\ 
    $R$ DMFT (J=0.7 eV) & 4.9 & 2.8 \\
    $R$ DMFT (J=1.0 eV) & 3.3 & 1.7 \\
    \hline \hline
  \end{tabular}
\end{center}
\caption{Parameters $P_0$ and $R$ of Eq.$\:$\ref{Pofd} resulting in the best linear fit to orbital polarization as a function of octahedral distortion $d$ for each Ni site and bulk. GGA and GGA+DMFT with $J$=0.7 and 1 eV are compared.}
\label{linearfit}
\end{table}

We now turn to the superlattice. Here the physics is richer.  Fig.~\ref{fig:pol} shows a pronounced difference in polarization  between A and B Ni ions, much larger changes across zero strain, and much greater variation of slopes.   From Fig.~\ref{fig:bondlength} we see also a greater richness of structural effects. The Ni-A and Ni-B sites respond differently to strain:   the distortion of the inner (Ni A) octahedron is greater than the distortion of the outer (Ni B) octahedron with the  A-B difference being greatest for compressive strain and becoming very small for large tensile strain.  

The primary reason for the difference in distortion is that the bonding between the apical O and the Al is weaker than the bonding between the apical O and the Ni B, so that the Al layer in effect provides a steric hinderance which prevents the O from coming too close to the Al, thus inhibiting the elongation of the Ni B-apical O bond length favored by compressive strain. Under tensile strain, a shorter Ni-apical O bond length is preferred, but the energy cost for increasing the Al-O bond length from its preferred value is much less than the cost of decreasing it, thus explaining the near equivalence of the structural distortions of the Ni-A and Ni-B under tensile strain. 

\begin{figure}[!htbp]
\includegraphics[width=0.85\columnwidth]{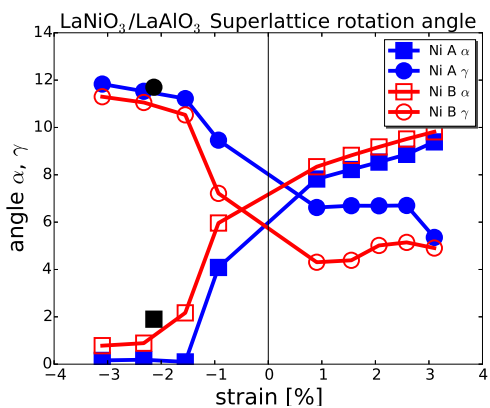}
\caption{(Color online) 
The layer-resolved NiO$_6$ octahedral rotational angles: $\alpha$, out-of-plane rotation around $x$ axis, (square dots) and $\gamma$, in-plane rotation around $z$ axis (circular dots) for 4/4 LaNiO$_3$/LaAlO$_3$ superlattices.  Ni A (filled dots) and Ni B (empty dots) are shown. The black dots under compressive strain denote experimental values obtained for 4/4 LaNiO$_3$/LaAlO$_3$~\cite{APL.104.2014}. \label{fig:bondangle}}
\end{figure}

The octrahedral rotations are also different for tensile and compressive strain. The rotation pattern for the NiO$_6$ octahedra itself changes from  the $a^-a^-c^-$  pattern observed at all strains in bulk and for compressive strain in the superlattice to  $a^-a^-c^+$ for tensile strain.  (The rotation pattern found for the AlO$_6$ octahedra is always  $a^-a^-c^-$).  Fig.$\:$\ref{fig:bondangle}  shows that for compressive strain the dominant rotation is around the $z$ axis and the out of plane rotation is negligible  $\gamma \gg \alpha \sim 0$  as is the case for the bulk materials. However, for  tensile strain both rotations occur, with rotations around the  in-plane somewhat larger but rotations about the z axis are not negligible.   The non-negligible  $\alpha$ found for tensile strain arises from a transverse motion of the oxygen which makes it easier for the system to reach a compromise between  the preferred Al-O and Ni-O in-plane bond lengths. 
With this information in hand we turn to the dependence of orbital polarization on octahedral distortion shown in Fig.~\ref{fig:pollcla}.  For the Ni-A site (panel a) both the superlattice orbital polarization $P$ (points) and the dependence of $P$ on the octahedral distortion ($R$) under tensile strain are very similar to that found for the bulk materials (lines); 
however for compressive strain the absolute $P$ values in the superlattice are rather larger than bulk while the effect of interactions (DMFT) is to strongly reduce $P$ toward bulk values.  The DMFT effect on the slope $R$ under compressive strain is smaller and indeed of opposite sign, leading for $J=0.7$eV to a slightly  larger $R$ than in GGA.

The most important effect apparent in the Ni-B site results (Fig.~\ref{fig:pollcla} (b)) is a large positive offset relative to the bulk calculation. Both for tensile and for compressive strain, and for all methods, the $P_0$ values are much larger in the superlattice Ni-B than in bulk or than for the Ni A-site. This difference in $P_0$ is the quantum confinement effect: the barrier imposed by the AlO$_2$ layer has the effect of making the apical oxygen hybridize more strongly with the Ni B-site, thus raising the energy of the frontier  antibonding Ni-O state and thereby depopulating the $3z^2-r^2$ orbital. The effect is strongest in the GGA calculation, and is reduced as $J$ is increased. 

Turning now to the variation with octahedral distortion, we see that for tensile strain the fitted slope $R$  is systematically smaller for the Ni-B site than for the Ni-A site or for the bulk calculation, and the actual data display a tendency to saturation. Conversely, for compressive strain the Ni B-site $R$ is systematically larger than the A-site or bulk values. These observations indicate a strong coupling between quantum confinement and structural effects.  For the B-site under tensile strain, the effect of the strain-induced decrease in apical Ni-O bond length is less significant because the O is already strongly bonded to the Ni and the quantum confinement effect is strong; this explains the decreased $R$ value compared to the A-site or bulk. For compressive strain, the tendency to increase the Ni-O distance weakens the Ni-O bond making the quantum confinement effect less important and conversely the octahedral distortion effect more influential; these effects explains the larger $R$ value under compressive strain. 
 
\section{Analysis: many-body effects \label{manybody}}

\begin{figure}[!htbp]
\includegraphics[width=0.85\columnwidth]{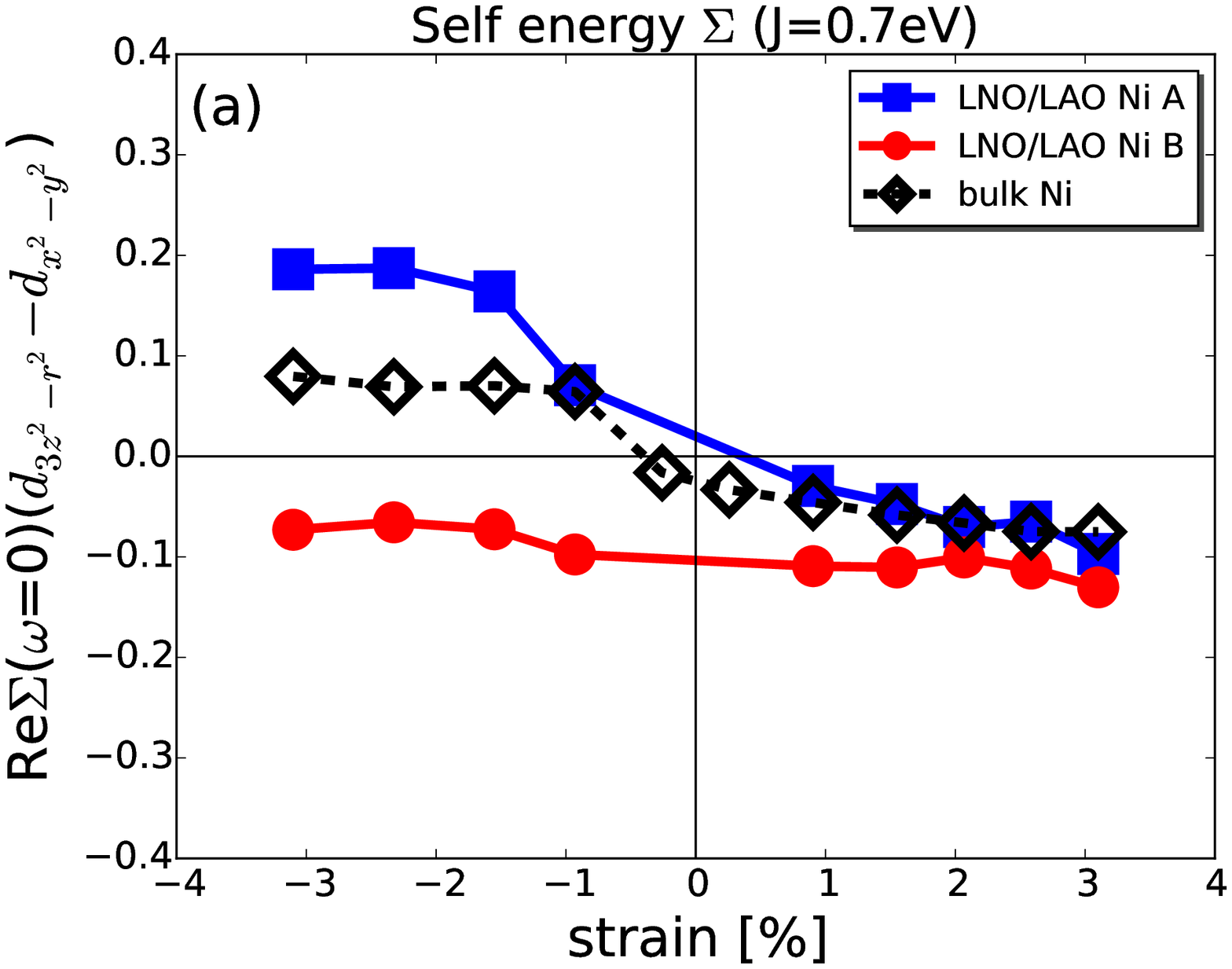}
\includegraphics[width=0.85\columnwidth]{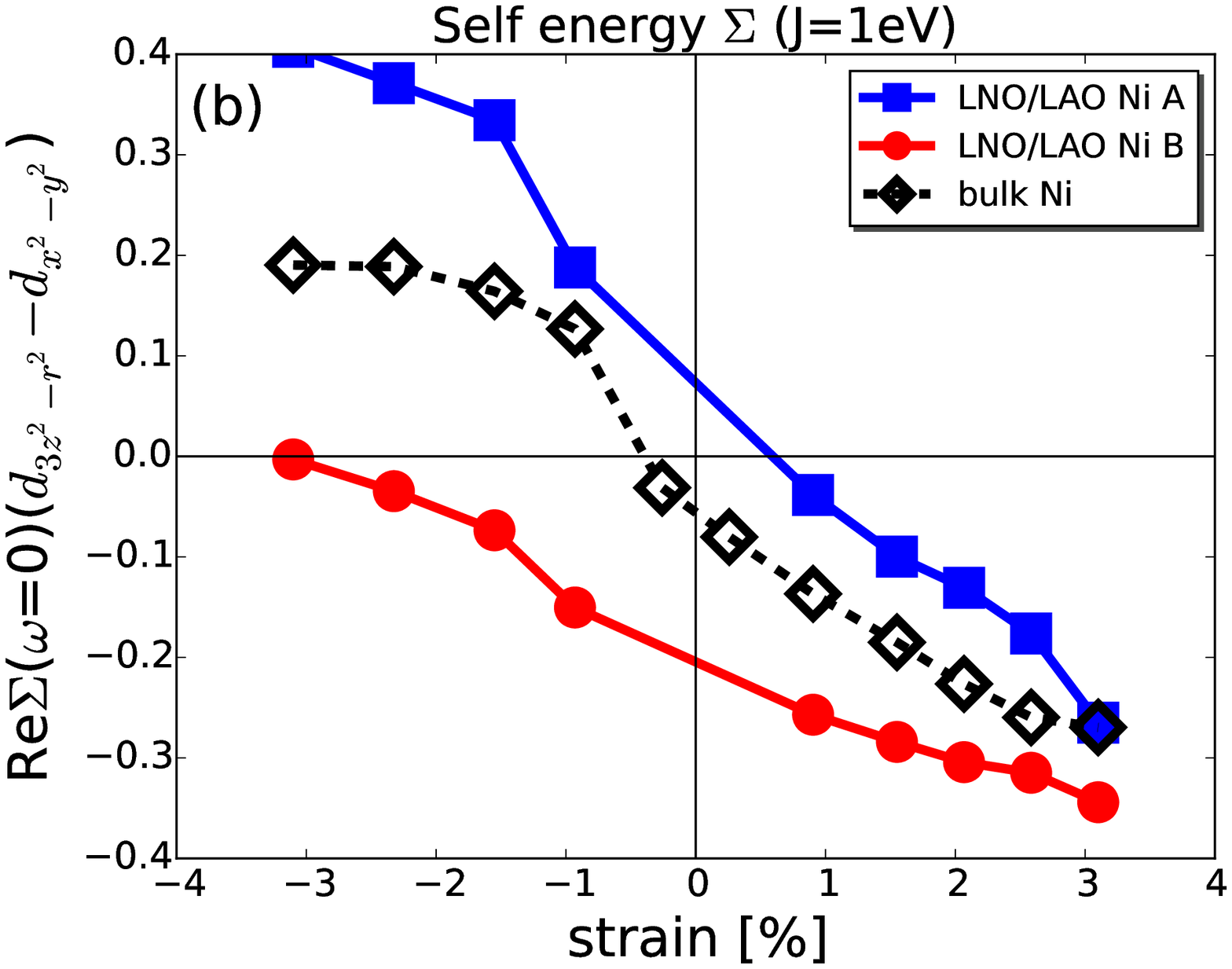}
\caption{(Color online) 
The difference of the real part of the self energy $\Sigma_d(\omega=0)$ between the $d_{3z^2-r^2}$ and the $d_{x^2-y^2}$ orbitals for the 4/4 LaNiO$_3$/LaAlO$_3$ superlattice (filled symbols and solid lines) and for strained bulk LaNiO$_3$ (open symbols and dashed line) for $U$=5eV and $J$-values indicated.\label{fig:Sig}}
\end{figure}

The effect of correlations on the orbital polarization is contained in the real parts of the electron self energies. The self energy is a $2\times 2$ matrix. In the DMFT approximation used here it is site-local  and for a given Ni site it is approximately diagonal  in the orbital  basis aligned with the axes of the NiO$_6$ octahedron. In this basis it has two components, corresponding to the $d_{x^2-y^2}$ and $d_{3z^2-r^2}$ orbitals.  In the GGA+DMFT approach used here the double counting term is orbital-independent, so  the difference between the real parts of the two diagonal components of the  self energy provides a many-body correction to the difference in bare energy levels. A positive sign of $\Sigma_{d_{3z^2-r^2}}-\Sigma_{d_{x^2-y^2}}$ means that many-body effects shift the $3z^2-r^2$ level up in energy relative to the $x^2-y^2$ level, thus increasing $P$. Fig.$\:$\ref{fig:Sig} presents the zero frequency limit of the self energy difference, for the superlattice and the bulk system.

Consider first the results for the bulk material. We see that the self energy difference is positive (acts to increase $P$) for compressive strain (except near zero strain where GGA $P$ starts to change the sign) and negative (acts to decrease $P$) for tensile strain. We see that for both signs of strain, the magnitude of the self energy is twice as large for $J=1$eV as it is for $J=0.7$eV, and that the self energy has more strain dependence for tensile strain than for compressive strain. 
The difference in magnitude of the self energy $\Sigma$ difference across $d=0$ causes the decrease in $P_0$ between compressive and tensile strain. A dependence of the {\em magnitude} of $\delta \Sigma$ on strain provides the renormalization of $R$. We see that for the bulk material, the change of the magnitude of $\Sigma$ is much larger than that of the $\delta \Sigma$ dependence on strain and this $\delta \Sigma$ dependence is larger for tensile than for compressive strain.  These results explain the strong reduction of $P_0$ compared to $R$ and the difference in renormalization of $R$ reported above.

We now turn to the superlattice. We see that generically the sign of the self energy is such that the many-body effects act to drive the polarization towards zero for both Ni sites regardless of strain or quantum confinement. Thus for the B-site the self energy difference is always negative (decreases $P$) because quantum confinement effects produce positive GGA $P$ values for all strain values. The Ni A self energy changes sign because the sign of $P$ also changes. The sign change does not take place at exactly the same strain value in both quantities because $P$ is determined by an average of $Re \Sigma$ over a range of frequencies. 

To further confirm the consistent behavior of orbital polarization for the ultra-thin limit, we also computed the 1/1 LaNiO$_3$/LaAlO$_3$ superlattice (which should exhibit stronger quantum confinement effects since the Ni site is bounded on two sides by the insulator) at $U$=5eV and $J$=0.7eV for the most compressive strain (-3.1\%) and the most tensile strain (3.1\%). Under compressive strain, Ni $P$ computed using GGA is -16.52\% while GGA+DMFT produces $P=7.4\%$. The large offset and the sign change are very similar to that seen in the B-site of the  4/4 superlattice, confirming that the effect is related to quantum confinement. Under tensile strain, GGA $P$ is 23.63\% and GGA+DMFT reduces the $P$ value to 15.27\%, a fractional reduction similar to that found for the B-site in the 4/4 superlattice, confirming that the quantum confinement effects are more important in this case.

\section{Conclusion\label{Conclusion}}

This paper presents a theoretical study of the layer-resolved orbital polarization of strained 4/4 LaNiO$_3$/LaAlO$_3$ perovskite superlattices. We used spin polarized GGA to obtain relaxed structures and paramagnetic GGA+DMFT to account for correlations on the Ni sites. Our calculations introduce two kinds of external symmetry breaking: $c$-axis quantum confinement associated with the insulating spacer layers and lattice strain. We further analyze these perturbations in terms of the resulting octahedral symmetry breaking caused by structural relaxations that lead to a difference between the apical ($c$-axis) Ni-O bond length $l_c$ and the in-plane bond length $l_a$, as well as rotational and tilting distortions of NiO$_6$ octahedra.  By comparing many-body and pure GGA calculations, as well as superlattice and strained bulk calculations, we are able to separate the effects. 

The results presented here indicate that strain affects orbital polarization in two ways: it deforms the NiO$_6$ octahedra, thereby explicitly leading to a  splitting of the two Ni $e_g$ states, and it changes the type of octahedral rotation pattern observed for tensile versus compressive strain. It is useful to express the polarization as the sum of a term proportional to the octahedral distortion of a NiO$_6$ octahedron and a residual arising from quantum confinement and octahedral rotation effects (see Table~\ref{linearfit}). While in strained bulk LaNiO$_3$ the change in rotation angles has only a small effect on orbital polarization, in the superlattice the effect is larger. We further find that proximity to the insulating AlO$_2$ layer has a dramatic effect on the polarization. This quantum confinement effect is at least as important as the strain effects, but is very local, affecting the outer-layer Ni B site substantially and the inner-layer Ni A site hardly at all. Finally, we note that quantum confinement and strain effects combine in interesting ways.  For tensile strain the superlattice Ni B exhibits a reduced $R$ value than bulk materials due to quantum confinement, while for compressive strain the $R$ value for Ni B can be larger since the octahedral distortion effect on the change of $P$ is more important.  

Our calculations reproduce the experiment~\cite{PRB.88.125124} semiquantitatively, yielding differences between the polarizations of the Ni A and B sites with about the correct order of magnitude and with a strain dependence of the correct order of magnitude. GGA+DMFT is clearly an improvement over pure GGA calculations. We demonstrate that the results have some sensitivity to the value of the on-site interaction $J$, and the optimal value to descibe experiment lies somewhere between the values of 0.7 and 1.0 eV used in this study.

\section*{Acknowlegements}
We thank A. Georges and O. Peil for helpful discussions and pointing out an inaccuracy in the structural relaxation calculations in an earlier version of this paper. H. Park acknowledges support of the start-up funding from UIC and Argonne National Laboratory (by the US Department of Energy, Office of Science program). The work of AJM on this project was supported by the basic energy sciences program of the US Department of Energy under grant ER-046169. CAM and H. Park acknowledge support from FAME, one of six centers of STARnet, a Semiconductor Research Corporation program sponsored by MARCO and DARPA. We gratefully acknowledge the computing resources provided on Blues and/or Fusion, a high-performance computing cluster operated by the Laboratory Computing Resource Center at Argonne National Laboratory. We also acknowledge computational facilities provided via XSEDE resources through Grant No. TG-PHY130003. A.J.M. acknowledges the  warm hospitality and stimulating intellectual environment of the College de France, where part of the writing of this manuscript was completed.
\bibliography{main}

\end{document}